\documentclass[aps,prl,twocolumn,showpacs,groupedaddress,nofootinbib]{revtex4}  
\usepackage{graphicx}  
\usepackage{dcolumn}   
\usepackage{bm}        
\usepackage{amsmath}   
\usepackage{amssymb}   
\usepackage{url}
\usepackage{xspace}

\newcommand{\ptmiss}{\ensuremath{{p\kern-0.5em\slash}_{T}}\xspace}
\newcommand{\alpgen}{{\sc alpgen}\xspace}
\newcommand{\pythia}{{\sc pythia}\xspace}
\newcommand{\vecbos}{{\sc vecbos}\xspace}
\newcommand{\geant}{{\sc geant}\xspace}

\begin{document}
\hspace{5.2in} \mbox{FERMILAB-PUB-08-242-E}

\title{Precise measurement of the top quark mass from lepton$+$jets events}
%
\author{V.M.~Abazov$^{36}$}
\author{B.~Abbott$^{75}$}
\author{M.~Abolins$^{65}$}
\author{B.S.~Acharya$^{29}$}
\author{M.~Adams$^{51}$}
\author{T.~Adams$^{49}$}
\author{E.~Aguilo$^{6}$}
\author{M.~Ahsan$^{59}$}
\author{G.D.~Alexeev$^{36}$}
\author{G.~Alkhazov$^{40}$}
% ...

\affiliation{\vspace{0.1 in}(The D\O\ Collaboration)\vspace{0.1 in}}
% ...
\affiliation{$^{6}$University of Alberta, Edmonton, Alberta, Canada,
                Simon Fraser University, Burnaby, British Columbia, Canada,
                York University, Toronto, Ontario, Canada, and
                McGill University, Montreal, Quebec, Canada}
% ...
\affiliation{$^{29}$Tata Institute of Fundamental Research, Mumbai, India}
% ...
\affiliation{$^{36}$Joint Institute for Nuclear Research, Dubna, Russia}
% ...
\affiliation{$^{40}$Petersburg Nuclear Physics Institute,
                St. Petersburg, Russia}
% ...
\affiliation{$^{49}$Florida State University, Tallahassee, Florida 32306, USA}
% ...
\affiliation{$^{51}$University of Illinois at Chicago,
                Chicago, Illinois 60607, USA}
% ...
\affiliation{$^{59}$Kansas State University, Manhattan, Kansas 66506, USA}
% ...
\affiliation{$^{65}$Michigan State University,
                East Lansing, Michigan 48824, USA}
% ...
\affiliation{$^{75}$University of Oklahoma, Norman, Oklahoma 73019, USA}
% ...
\date{September 16, 2008}

\begin{abstract}
We measure the mass of the top quark using top quark pair candidate events in
the lepton+jets channel from data corresponding to 1 fb$^{-1}$ of integrated
luminosity collected by the D0 experiment at the Fermilab Tevatron collider. We
use a likelihood technique that reduces the jet energy scale uncertainty by
combining an \emph{in-situ} jet energy calibration with the independent
constraint on the jet energy scale (JES) from the calibration derived using
photon+jets and dijet samples. We find the mass of the top quark to be
$171.5\pm1.8(\mbox{stat.+JES})\pm1.1(\mbox{syst.})$ GeV. 
\end{abstract}

\maketitle

Since the discovery of the top quark in 1995~\citep{topdiscovery}, a substantial
effort has gone into measuring and understanding its properties. Its large mass
suggests a unique role in the mechanism of electroweak symmetry breaking.
Through radiative corrections, a precise measurement of the top quark mass,
together with that of the $W$ boson, allows indirect constraints to be placed on
the mass of the standard model Higgs boson~\citep{lepewwg}. A precise knowledge
of the top quark mass could also provide a useful constraint to possible
extensions of the standard model. It is therefore of great importance to
continue improving measurements of the top quark
mass~\citep{topmassd0prd,topmasscdf}.

In this Letter, we present the most precise single measurement of the top quark
mass from Run~II of the Fermilab Tevatron collider. It uses a matrix element
(ME) method with an \emph{in-situ} jet energy calibration based on a
global factor used to scale all jet energies and thereby the invariant
mass of the hadronic $W$ boson~\citep{topmassd0prd,topmassd0nature}. This mass
is constrained to the well known value of 80.4 GeV through the Breit-Wigner
function for the hadronic $W$ boson in the ME for $t\bar{t}$ production. The jet
energy scale is further constrained to the standard scale derived from
photon+jets and dijet samples within its uncertainties through the use of a
prior probability distribution. This analysis is based on data collected by the
D0 detector~\citep{d0nim} from April 2002 to February 2006 comprising 1
fb$^{-1}$ of integrated luminosity from $p\overline{p}$ collisions at
$\sqrt{s}=1.96$~TeV.

The top quark is assumed to always decay into a $W$ boson and a $b$ quark
producing a $W^{+}W^{-}b\overline{b}$ final state from $t\overline{t}$
production. This analysis is based on the lepton$+$jets channel with one $W$
boson decaying via $W\rightarrow\ell\nu$ and the other via $W\rightarrow
q\overline{q}^{\prime}$. This channel is characterized by a lepton with large
transverse momentum ($p_{T}$), large momentum imbalance due to the undetected
neutrino (\ptmiss), and four high-$p_{T}$ jets. Events are selected for this
analysis by requiring exactly one isolated electron (muon) with $p_{T}>20$~GeV
and $\left|\eta\right|<1.1$ ($\left|\eta\right|<2$), $\ptmiss>20$~GeV, and
exactly four jets with $p_{T}>20$~GeV and $\left|\eta\right|<2.5$, where the
pseudorapidity $\eta=-\ln\left[\tan(\theta/2)\right]$, and $\theta$ is the polar
angle with respect to the proton beam direction.  Multijet background, typically
originating from lepton or jet energy mismeasurements, is further suppressed by
requiring the lepton direction and \ptmiss vector to be separated in azimuth.
Jets are reconstructed using a cone algorithm~\citep{run2jets} with radius
$R=\sqrt{(\Delta y)^{2}+(\Delta\phi)^{2}}=0.5$ where the $y$ is the rapidity.
Jet energies are corrected to the particle level using corrections derived from
photon+jet and dijet samples. Jets containing a muon are assumed to originate
from semileptonic $b$ quark decays and corrected by the muon momentum and
average neutrino energy. At least one jet is required to be tagged by a
neural-network based algorithm~\citep{scanlon} as a $b$-jet candidate. The
tagging efficiency for $b$ jets is $\sim50$\% with a misidentification rate of
$\sim1$\% from light jets. A total of 220 events, split equally between the
electron and muon channels, satisfying these criteria is selected.

The top quark mass is determined from the data sample with a likelihood method
based on per-event probability densities (p.d.'s) constructed from the MEs of
the processes contributing to the observed events. Assuming only two processes,
$t\overline{t}$ and $W$+jets production, the p.d. to observe an event with
measured variables $x$ is
\[
P_{\textrm{evt}}=A(x)\left[fP_{\textrm{sig}}(x;m_{t},k_{\textrm{jes}})+(1-f)P_{\textrm{bkg}}(x;k_{\textrm{jes}})\right],
\]
where the top quark mass $m_{t}$, jet energy scale factor $k_{\textrm{jes}}$
dividing the energies of all jets, and observed signal fraction $f$ are the
parameters to determine from data. $P_{\textrm{sig}}$ and $P_{\textrm{bkg}}$
are, respectively, p.d.'s for $t\overline{t}$ and $W$+jets production. Multijet
events satisfy $P_{\textrm{bkg}}\gg P_{\textrm{sig}}$ and are also represented
by $P_{\textrm{bkg}}$. $A(x)$ is a function only of $x$ and accounts for the
geometrical acceptance and efficiencies.

$P_{\textrm{sig}}$ and $P_{\textrm{bkg}}$ are calculated by integrating over all
possible parton states leading to the measured set $x$. In addition to the
partonic final state described by the variables $y$, these states include the
initial state partons carrying momenta $q_{1}$ and $q_{2}$ in the colliding $p$
and $\overline{p}$. The integration involves a convolution of the partonic
differential cross section $d\sigma(y;m_{t})$ with the p.d.'s for
the initial state partons $f(q_{i})$ and the transfer function
$W(y,x;k_{\textrm{jes}})$:
\[
P_{\textrm{sig}}=\frac{1}{N}\int{\displaystyle \sum}d\sigma(y;m_{t})dq_{1}dq_{2}f(q_{1})f(q_{2})W(y,x;k_{\textrm{jes}}),
\]
where the sum runs over all possible initial state parton flavor combinations.
$f(q_{i})$ includes parton density functions (PDFs) for finding a parton of a
given flavor and longitudinal momentum fraction in the $p$ or $\overline{p}$
(CTEQ6L1~\citep{cteq}) and parameterizations of the p.d.'s for the transverse
components of $q_{i}$ derived from \pythia~\citep{pythia}. Jet fragmentation
effects and experimental resolution are taken into account by
$W(y,x;k_{\textrm{jes}})$, representing the p.d. for the measured set $x$ to
have arisen from the partonic set $y$. The normalization factor $N$, defined as
the expected observed cross section for a given
($m_{\textrm{t}}$,$k_{\textrm{jes}}$), ensures $A(x)P_{\textrm{sig}}$ (and
ultimately $P_{\textrm{evt}}$) is normalized to unity.

The differential cross section term in $P_{\textrm{sig}}$ is calculated using
the leading order ME for $q\overline{q}\rightarrow t\overline{t}$. After all
energy and momentum constraints are taken into account, this term is integrated
over the energy associated with one of the quarks from the hadronic $W$ boson
decay, the masses of the two $W$ bosons and two top quarks, and the energy
($1/p_{T}$) of the electron (muon). It is summed over 24 possible jet-parton
assignments each carrying a $b$-jet tagging weight~\citep{def:wgt_btg} and over
the neutrino solutions. $W(y,x;k_{\textrm{jes}})$ is the product of five terms
for the four jets and one charged lepton. The jet terms are parameterized in
terms of jet energy with a function involving the sum of two Gaussians whose
parameters depend linearly on parton energy. The term for the charged lepton is
parameterized as a Gaussian distribution in energy ($1/p_{T}$) for electrons
(muons). All parameters for $W(y,x;k_{\textrm{jes}})$ are derived using fully
simulated Monte Carlo (MC) events. The normalization cross section 
$\sigma_{\textrm{obs}}^{t\overline{t}}=\smallint A(x)P_{\textrm{sig}}dx=
\sigma^{t\overline{t}}(m_{t})\left\langle
A(m_{t},k_{\textrm{jes}})\right\rangle$ is calculated using the total cross
section corresponding to the ME used and the mean acceptance for events  whose
dependencies on $m_{t}$ and $k_{\textrm{jes}}$ are determined from MC events.

The differential cross section term in $P_{\textrm{bkg}}$ is calculated using
the $W+$4 jets MEs provided by \vecbos~\citep{vecbos}. The integration is
performed over the energies of the four partons producing the jets, the $W$
boson mass, and the energy ($1/p_{T}$) of the electron (muon) summing over 24
possible jet-parton assignments and two neutrino solutions. The transverse
momenta of the initial state partons are assumed to be zero.

$P_{\textrm{sig}}$ and $P_{\textrm{bkg}}$ are calculated using MC techniques on
a grid in ($m_{t}$,$k_{\textrm{jes}}$) having spacings of
1.5 GeV and 0.015, respectively. At each grid point, a likelihood function,
$L(x;m_{t},k_{\textrm{jes}},f)$, is constructed from the product of the
individual event p.d.'s ($P_{\textrm{evt}}$) and $f$ is determined by
minimizing $-\ln L$ at that point. $L(x;m_{t},k_{\textrm{jes}})$ is then
projected onto the $m_{t}$ and $k_{\textrm{jes}}$ axes according to
$L(x;m_{t})={\displaystyle \smallint
L(x;m_{t},k_{\textrm{jes}})G(k_{\textrm{jes}})dk_{\textrm{jes}}}$ and
$L(x;k_{\textrm{jes}})=\smallint L(x;m_{t},k_{\textrm{jes}})dm_{t}$. The prior
$G(k_{\textrm{jes}})$ is a Gaussian function centered at $k_{\textrm{jes}}=1$
with width $\sigma=0.019$ determined from the photon+jets and dijet samples used
in the standard jet energy scale calibration. Best estimates of $m_{t}$ and
$k_{\textrm{jes}}$ and their uncertainties are then extracted from
the mean and rms of $L(x;m_{t})$ and $L(x;k_{\textrm{jes}})$.
\begin{figure}
\begin{centering}
\includegraphics[width=1\columnwidth]{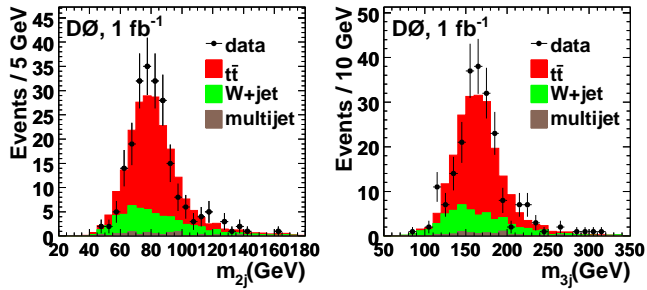} 
\par\end{centering}
\caption{\label{fig:datamc}Comparison between data and MC 2-jet ($m_{\textrm{2j}}$)
and 3-jet ($m_{\textrm{3j}}$) invariant mass distributions.}
\end{figure}

The measurement technique described above is calibrated using MC events produced
with the \alpgen event generator~\citep{alpgen} employing \pythia for parton
showering and hadronization and implementing the MLM matching
scheme~\citep{mlm}. All generated events are processed by a full
\geant~\citep{geant} detector simulation followed by the same reconstruction and
analysis programs used on data. Fig.~\ref{fig:datamc} shows comparisons of the
2-jet and 3-jet invariant mass distributions between data and MC using
$t\overline{t}$ events with a true top quark mass ($m_{t}^{\textrm{true}}$) of
170 GeV. These are calculated using jets assigned as the decay products of the
top quark and $W$ boson from the hadronic branch in the permutation with the
largest weight (defined as the product of $P_{\textrm{sig}}$ and the $b$-jet
tagging weight) around the peak of $L(x;m_{t},k_{\textrm{jes}})$. MC
distributions are normalized to data distributions with $f=0.74$ determined from
data. The background includes simulated $W$+jets events and data events selected
from a multijet enriched sample.  The latter comprises 12\% of the total
background based on estimates from data. The estimated number of $t\bar{t}$
events ($e+\textrm{jets}$: $91\pm9$, $\mu+\textrm{jets}$: $71\pm8$) agrees with
the expectation ($e+\textrm{jets}$: $89\pm6$, $\mu+\textrm{jets}$: $73\pm5$).
\begin{figure}
\begin{centering}
\includegraphics[width=0.5\columnwidth]{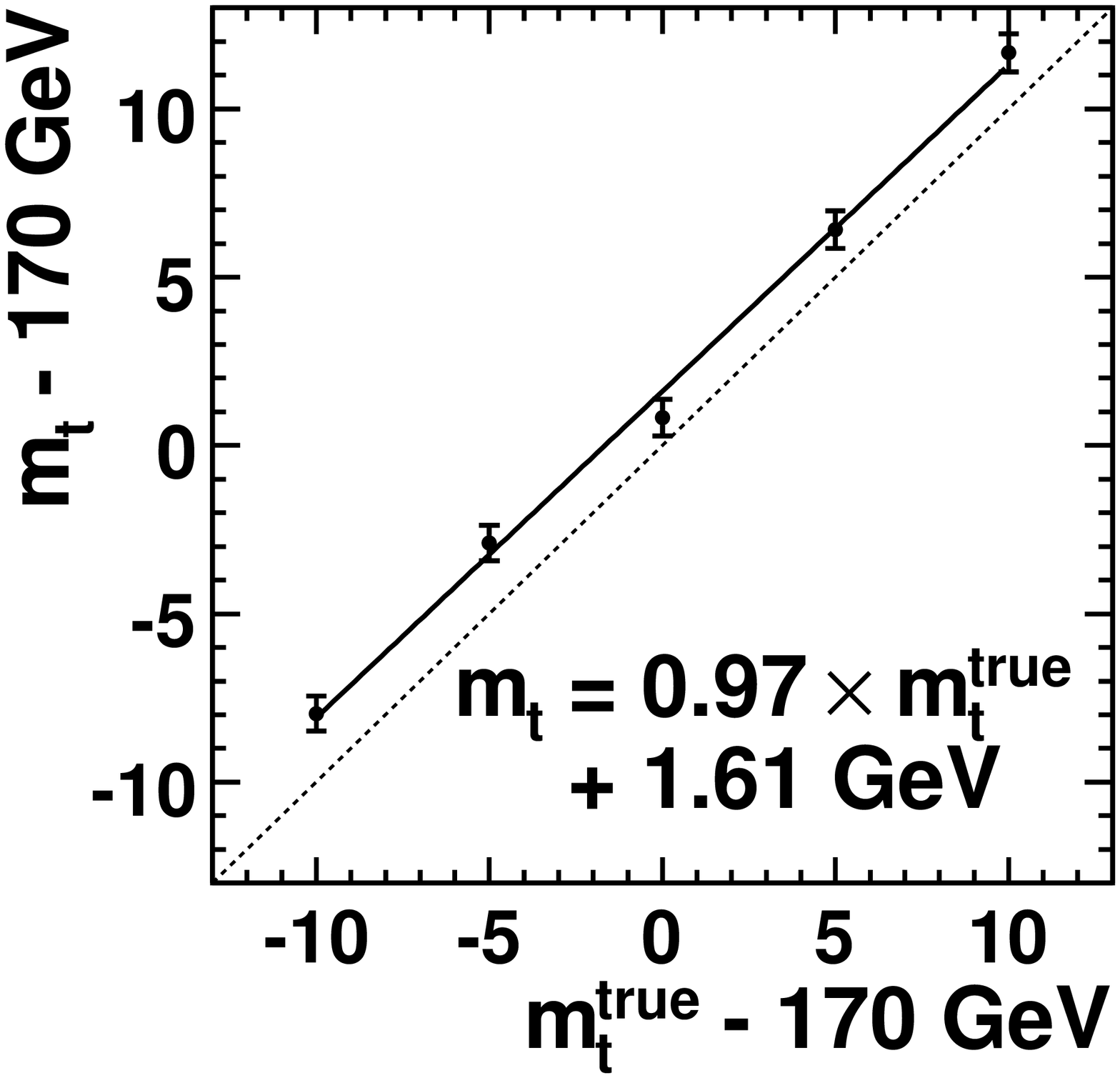}\includegraphics[width=0.5\columnwidth]{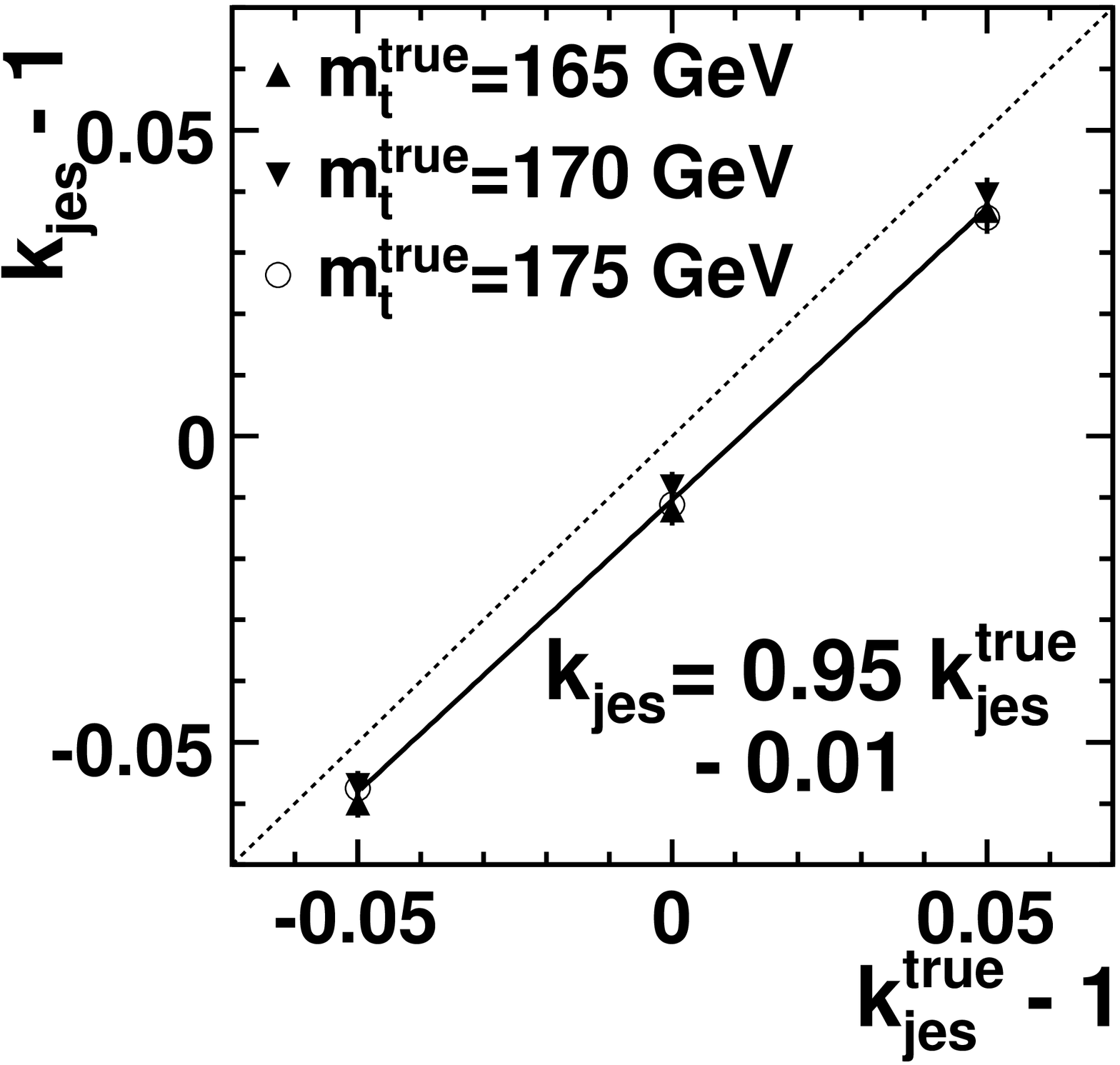} 
\par\end{centering}
\caption{\label{fig:calib}Mean values of $m_{t}$ and $k_{\textrm{jes}}$
from ensemble tests versus true values parameterized by straight lines.
Dashed lines represent identical fitted and true values.}
\end{figure}

Five $t\overline{t}$ MC samples are generated with $m_{t}^{\textrm{true}}=160$,
165, 170, 175, and 180 GeV, with six more produced from three of these by
scaling all jet energies by $\pm5$\%. $P_{\textrm{sig}}$ and $P_{\textrm{bkg}}$
are calculated for these events from which  pseudo-experiments fixed to the
number of data events are randomly drawn with a signal fraction fluctuated
according to a binomial distribution around that determined from data. The mean
values of $m_{t}$ and $k_{\textrm{jes}}$ for 1000 pseudo-experiments are shown
in Fig.~\ref{fig:calib} as functions of the true values and fitted to a straight
line. The average widths of the $m_{t}$ and $k_{\textrm{jes}}$ pull
distributions are 1.0 and 1.1, respectively. The pull is defined as the
deviation of a measurement from the mean of all measurements divided by the
uncertainty of the measurement per pseudo-experiment. The measured uncertainties
in data are corrected by the deviation of the average pull width from 1.0.

\begin{figure}
\includegraphics[width=0.5\columnwidth]{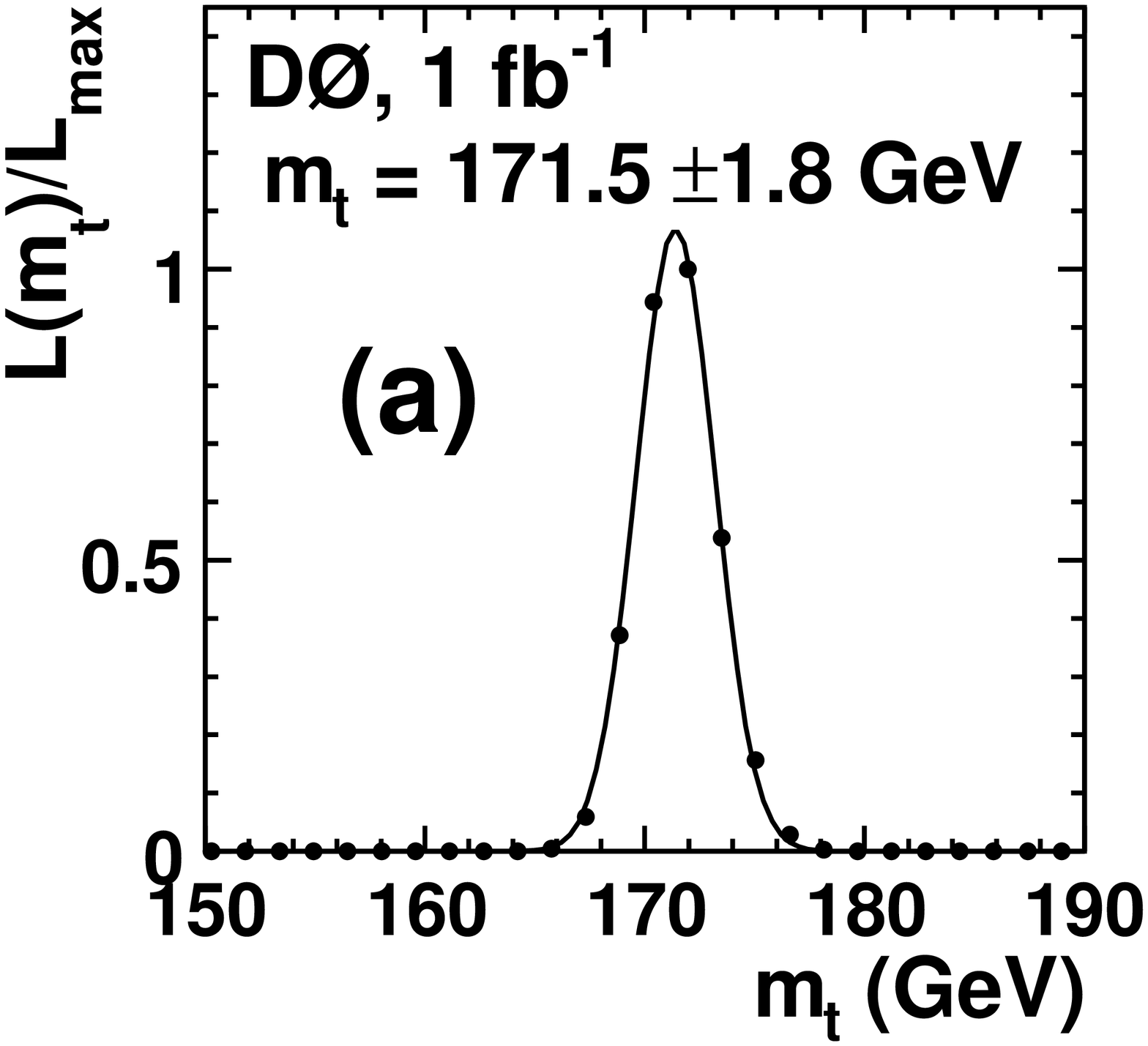}\includegraphics[width=0.5\columnwidth]{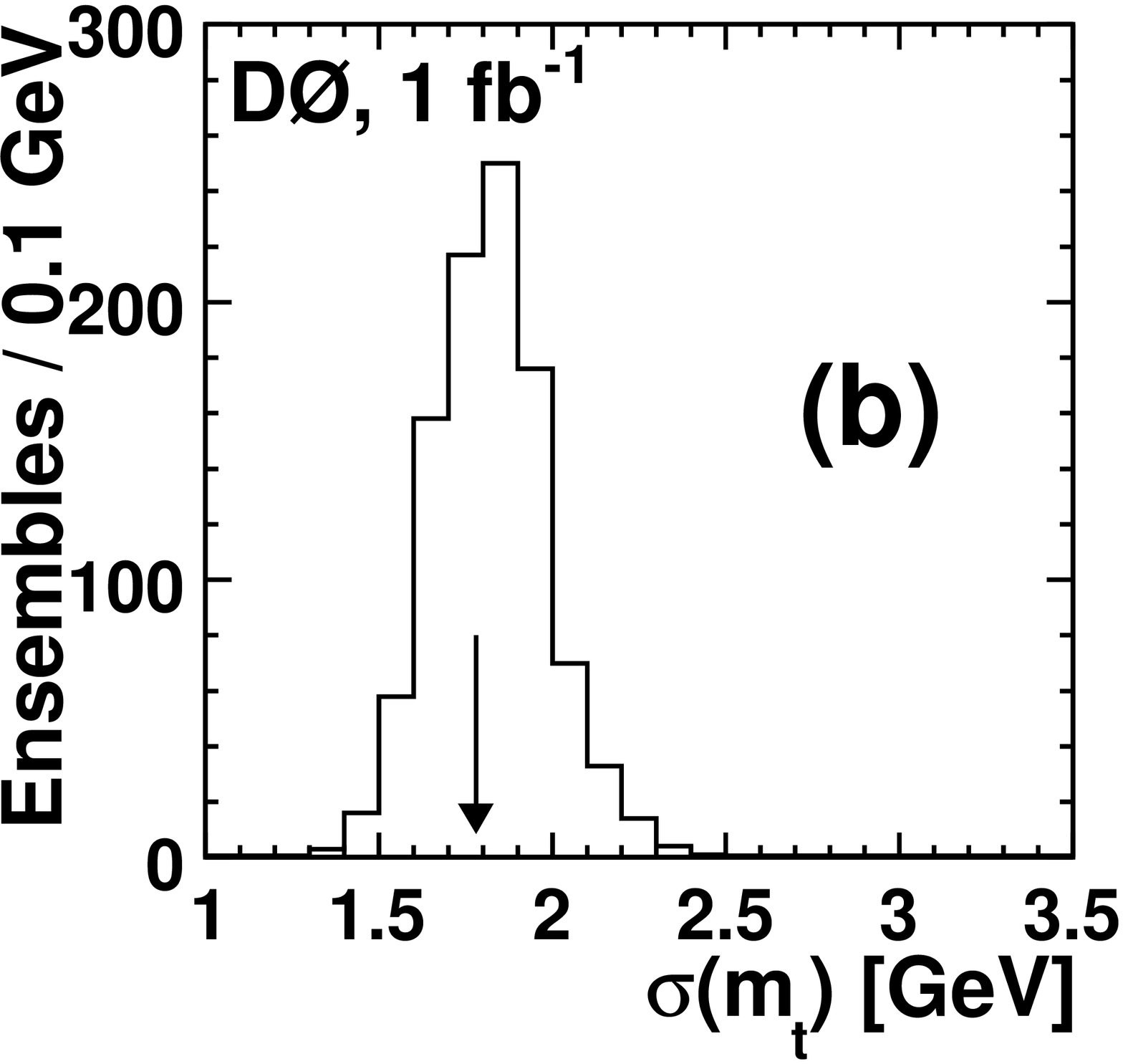}
\caption{\label{fig:results}(a) Projection of data likelihood onto the $m_{t}$
axis with best estimate shown. (b) Expected uncertainty distribution for $m_{t}$ with measured
uncertainty indicated by the arrow.}
\end{figure}

$L(x;m_{t})$ and $L(x;k_{\textrm{jes}})$ for the selected data samples are
calibrated according to the parameterizations shown in Fig.~\ref{fig:calib}.
$L(x;m_{t})$ is shown in Fig.~\ref{fig:results}(a) with a measured
$m_{t}=171.5\pm1.8(\textrm{stat.+JES})$~GeV. The measured
$k_{\textrm{jes}}=1.030\pm0.017$ represents a 1.2~$\sigma$ shift from
$k_{\textrm{jes}}=1$ where $\sigma$ is the sum in quadrature of the width of
$G(k_{\textrm{jes}})$ and the uncertainty of the measured $k_{\textrm{jes}}$.
Fig.~\ref{fig:results}(b) compares the measured uncertainty for $m_{t}$ with the
expected uncertainty distribution from pseudo-experiments in MC assuming
$m_{t}^{\textrm{true}}=170$~GeV.
\begin{table}
\caption{\label{tab:syst}Summary of systematic uncertainties (symmetrized
based on the larger of the two values in each direction).}
\vspace{.25cm}
\centering
\begin{tabular}{lc}
\hline
\hline 
Source  & Uncertainty (GeV)\\
\hline
\multicolumn{2}{l}{\textit{Physics modeling:}}\\
\hspace{12pt}Signal modeling  & $\pm0.40$\\
\hspace{12pt}PDF uncertainty  & $\pm0.14$\\
\hspace{12pt}Background modeling  & $\pm0.10$\\
\hspace{12pt}$b$-fragmentation  & $\pm0.03$\\
\multicolumn{2}{l}{\textit{Detector Modeling:}}\\
\hspace{12pt}$b$/light response ratio  & $\pm0.83$\\
\hspace{12pt}Jet identification and resolution  & $\pm0.26$\\
\hspace{12pt}Trigger  & $\pm0.19$\\
\hspace{12pt}Residual jet energy scale  & $\pm0.10$\\
\hspace{12pt}Muon resolution  & $\pm0.10$\\
\multicolumn{2}{l}{\textit{Method:}}\\
\hspace{12pt}MC calibration  & $\pm0.26$\\
\hspace{12pt}$b$-tagging efficiency  & $\pm0.15$\\
\hspace{12pt}Multijet contamination  & $\pm0.14$\\
\hspace{12pt}Signal contamination  & $\pm0.13$\\
\hspace{12pt}Signal fraction  & $\pm0.09$\\
\hline
Total  & $\pm1.07$\\
\hline
\hline
\end{tabular}
\end{table}

To verify the \emph{in-situ} jet energy calibration, we repeat the
analysis on data by fixing $k_{\textrm{jes}}$ to the measured value and
removing the $W$ boson mass constraint, replacing
$L(x;m_{t},k_{\textrm{jes}},f)$ with $L(x;m_{t},m_{W},f)$. $P_{\textrm{sig}}$
and $P_{\textrm{bkg}}$ are now calculated on a grid in ($m_{t}$,$m_{W}$) having
spacings of 1.5 GeV and 1 GeV, respectively. $L(x;m_{t})$ and $L(x;m_{W})$ are
calculated in the same way as for the grid in ($m_{t}$,$k_{\textrm{jes}}$)
except that no prior probability distribution is used for $L(x;m_{t})$. We find
$m_{W}=80.3\pm1.0$ GeV which is consistent with the constraint of $80.4$ GeV.

Systematic uncertainties are evaluated for three categories. The first category
involves the modeling of MC $t\overline{t}$ and $W+$jets events and includes
uncertainties in the modeling of extra jets due to radiation in $t\overline{t}$
events, the distribution shapes and the heavy flavor fraction in $W+$jets
events, $b$ fragmentation, and the PDFs used in generating events. The second
category is associated with the simulation of detector response and includes
possible effects due to the energy and $|\eta|$ dependence of the jet energy
scale unaccounted for by the \emph{in-situ} calibration, uncertainties in the
modeling of the relative calorimeter response to $b$ and light quark jets, and
uncertainties associated with the simulation of jet energy resolution and
reconstruction efficiency, muon $p_{T}$ resolution, and trigger efficiency. The
third category is related to assumptions made in the method and uncertainties in
the calibration and includes possible effects due to the exclusion of multijet
events and non-lepton+jets $t\overline{t}$ events from the calibration
procedure, uncertainties in the signal fraction used in ensemble tests, and
uncertainties associated with the parameters defining the calibration curve.
Contributions from all these sources are summarized in Table~\ref{tab:syst} and
sum in quadrature to $\pm1.1$ GeV.

The leading sources of uncertainty in Table \ref{tab:syst} are those
associated with the $b$/light response ratio and signal modeling. The first of
these is evaluated by estimating the possible difference in this ratio between
data and MC and scaling the energies of all jets matched to $b$ quarks in a MC
$t\bar{t}$ sample by this amount. The analysis is repeated for this sample and
the difference in $m_{t}$ from that of the unscaled sample taken as the
uncertainty. The uncertainty associated with the modeling of additional jets in
$t\overline{t}$ events is evaluated using both data and MC samples. Using MC
$t\overline{t}$ events, the fraction of $t\overline{t}$ signal events with
$\geq5$ jets is varied such that the ratio of $4$-jet to $\geq5$-jet events in
MC matches that in data including its uncertainties. The difference in the
resulting $m_{t}$ from that of the default sample is taken as the
uncertainty. Using data, this is done through ensemble tests in which a fixed
number of $\geq5$-jet events not used in the measurement are randomly drawn for
each experiment and combined with the default sample of $4$-jet
events. The ensemble tests are repeated for different fractions of $\geq5$-jet
events constituting up to 30\% of each experiment. $m_{t}$ for the
default sample is compared with the mean from each ensemble test and the
largest difference taken as the systematic uncertainty. Both procedures yield
consistent estimates for this systematic uncertainty.

In summary, we present a measurement of the top quark mass using $t\overline{t}$
lepton+jets events from 1 fb$^{-1}$ of data collected by the D0 experiment.
Using a ME technique combining an \emph{in-situ} calibration of the jet energy
scale with the calibration based on the photon+jets and dijet samples gives us
\[
m_{t}=171.5\pm1.8(\textrm{stat.+JES})\pm1.1(\textrm{syst.}) \textrm{ GeV},
\]
representing the most precise single measurement to date.

%
We thank the staffs at Fermilab and collaborating institutions, 
and acknowledge support from the 
DOE and NSF (USA);
CEA and CNRS/IN2P3 (France);
FASI, Rosatom and RFBR (Russia);
CNPq, FAPERJ, FAPESP and FUNDUNESP (Brazil);
DAE and DST (India);
Colciencias (Colombia);
CONACyT (Mexico);
KRF and KOSEF (Korea);
CONICET and UBACyT (Argentina);
FOM (The Netherlands);
STFC (United Kingdom);
MSMT and GACR (Czech Republic);
CRC Program, CFI, NSERC and WestGrid Project (Canada);
BMBF and DFG (Germany);
SFI (Ireland);
The Swedish Research Council (Sweden);
CAS and CNSF (China);
and the
Alexander von Humboldt Foundation (Germany).
%


\begin{thebibliography}{99}
%
\bibitem[a]{alton}
Visitor from Augustana College, Sioux Falls, SD, USA.
\bibitem[b]{burdin}
Visitor from The University of Liverpool, Liverpool, UK.
\bibitem[c]{podesta-lerma}
Visitor from ECFM, Universidad Autonoma de Sinaloa, Culiac\'an, Mexico.
\bibitem[d]{quadt,meyer,hensel,park}
Visitor from II. Physikalisches Institut, Georg-August-University,
  G{\"o}ttingen, Germany.
\bibitem[e]{voutilainen}
Visitor from Helsinki Institute of Physics, Helsinki, Finland.
\bibitem[f]{wenger}
Visitor from Universit{\"a}t Z{\"u}rich, Z{\"u}rich, Switzerland.
\bibitem[\ddag]{deceased}
Deceased.

%
\vskip 0.25cm

\bibitem{topdiscovery}F. Abe \emph{et al.} (CDF Collaboration), Phys.
Rev. Lett. \textbf{74}, 2626 (1995); S. Abachi \emph{et al.} (D0 Collaboration),
Phys. Rev. Lett. \textbf{74}, 2632 (1995).

\bibitem{lepewwg}Tevatron Electroweak Working Group, \url{http://tevewwg.fnal.gov}.

\bibitem{topmassd0prd}V. Abazov \emph{et al.} (D0 Collaboration),
Phys. Rev. D \textbf{74}, 092005 (2006).

\bibitem{topmasscdf}T. Aaltonen \emph{et al.} (CDF Collaboration),
Phys. Rev. Lett. \textbf{99}, 182002 (2007).

\bibitem{topmassd0nature}V. Abazov \emph{et al.} (D0 Collaboration),
Nature \textbf{429}, 638 (2004).

\bibitem{d0nim}V. Abazov \emph{et al.} (D0 Collaboration), Nucl.
Instrum. Methods Phys. Res. Sect. A \textbf{565}, 463 (2006).

\bibitem{run2jets}G.C. Blazey \emph{et al.}, arXiv:hep-ex/0005012
(2000).

\bibitem{scanlon}T. Scanlon, Ph.D. thesis, FERMILAB-THESIS-2006-43.

\bibitem{cteq}J. Pumplin \emph{et al.}, JHEP \textbf{0207}, 012 (2002).

\bibitem{pythia}T. Sj$\ddot{\mbox{o}}$strand \emph{et al.}, Comp.
Phys. Commun. \textbf{135}, 238 (2001).

\bibitem{def:wgt_btg}The weight for a $b$-tagged jet with a given $p_T$ and
$\eta$ under a parton flavor hypothesis $\alpha$(=$b$, $c$, light $q$ or gluon)
is given by the parameterization of the average tagging efficiency
$\epsilon_\alpha(p_T,\eta)$. Consequently, the weight for a jet not $b$-tagged
is 1-$\epsilon_\alpha(p_T,\eta)$. The event weight is defined as the product of
jet weights.

\bibitem{vecbos}F.A. Berends \emph{et al.}, Nucl. Phys. \textbf{B357},
32 (1991).

\bibitem{alpgen}M.L. Mangano \emph{et al.}, JHEP \textbf{307}, 001
(2003).

\bibitem{mlm}S. H$\ddot{\mbox{o}}$che \emph{et al.}, arXiv:hep-ph/0602031
(2006).

\bibitem{geant}R. Brun and F. Carminati, CERN Program Library Long Writeup
W5013, 1993 (unpublished). 

\end{thebibliography}
\end{document}